\documentclass[11pt]{article}
\usepackage{amsfonts}
\usepackage{amsmath}

\newcommand{\be}{\begin{equation}}
\newcommand{\ee}{\end{equation}}
\newcommand{\bea}{\begin{eqnarray}}
\newcommand{\eea}{\end{eqnarray}}

\makeatletter

\@addtoreset{equation}{section}
\def\section{\@startsection {section}{1}{\z@}{-3.5ex plus -1ex minus
 -.2ex}{2.3ex plus .2ex}{\large\bf\centering}}
\def\subsection{\@startsection{subsection}{2}{\z@}{-3.25ex plus -1ex minus -.2ex}{1.5ex plus .2ex}{\bf}}
\def\subsubsection{\@startsection{subsubsection}{3}{\z@}{-3.25ex plus -1ex minus -.2ex}{1.5ex plus .2ex}{\sl}}
\makeatother
\begin{document}

\baselineskip 18pt \parindent 12pt \parskip 12pt

\begin{titlepage}

\begin{center}
{\Large {\bf Quasideterminant solutions of the generalized Heisenberg magnet model}}\\\vspace{1in} {\large U. Saleem  \footnote{%
Tel No: +92-42-99231243, Fax No: +92-42-35856892\\e-mail:usaleem@physics.pu.edu.pk, usman\_physics@yahoo.com}  and M. Hassan   \footnote{%
mhassan@physics.pu.edu.pk } }\vspace{0.15in}

{{\it Department of Physics, University of the Punjab,\\
Quaid-e-Azam Campus, Lahore-54590, Pakistan.}}\\

\end{center}
\vspace{1cm}

\begin{abstract}
In this paper we present Darboux transformation for the
generalized Heisenberg magnet (GHM) model based on general linear
Lie group $GL(n)$ and construct multi-soliton solutions in terms
of quasideterminants. Further we relate the quasideterminant
multi-soliton solutions obtained by the means of Darboux
transformation with those of obtained by dressing method. We also
discuss the model based on the Lie group $SU(n)$ and obtain
explicit soliton solutions of the model based on $SU(2)$.
\end{abstract}
\vspace{1cm} PACS: 11.10.Nx, 02.30.Ik

\end{titlepage} 

\section{Introduction} \label{introduction}

During the past decades, there has been an increasing interest in
the study of classical and quantum integrability of Heisenberg
ferromagnet (HM) model \cite{Cherednik:1996wf}-\cite{GHM10}. The
Heisenberg ferromagnet (HM) model based on Hermitian symmetric
spaces has been studied in \cite{Symmetric1}-\cite{Symmetric4}.
The integrability of the HM model based on $SU(2)$ via inverse
scattering method is presented in \cite{GHM1}-\cite{GHM2} and its
$SU(n)$ generalization is studied in \cite{GHM3}. The
integrability of the GHM model based on the general linear Lie
group $GL(n)$ via Lax formalism has been investigated in
\cite{Cherednik:1996wf}. In this paper we present the Darboux
transformation of the GHM model based on general linear group
$GL(n)$ with Lie algebra $\verb"gl(n)"$ and calculate
multi-soliton solutions in term of quasideterminants. We also
establish the relation between the Darboux transformation and the
well-known  dressing method \cite{zakha}. In the last section, we
discuss the model based $SU(n)$ and calculate an explicit
expression of the single-soliton solution of the HM model based on
the Lie group $SU(2)$ using Darboux transformation.

The Hamiltonian of the GHM model is defined by
\cite{Cherednik:1996wf}
\begin{equation}
{\cal H}=%
\frac{1}{2}\mbox{Tr}\left(\left(\partial_{x}U\right)^{T}\left(\partial_{x}U\right)\right)
, \label{action}
\end{equation}%
with $"T"$ is transpose and $U(x,t)$ is a matrix-valued function
which takes values in the Lie algebra $\verb"gl(n)"$ of the
general linear group $GL(n)$. The corresponding equation of motion
can be expressed as
\begin{equation}
\partial_{t}U=\{ {\cal H},\partial_{x}U\}. \label{time evolution}
\end{equation}%
The above equation (\ref{time evolution}) can be written as%
\begin{equation}
\partial_{t}U=\left[U,\partial^{2}_{x}U\right], \label{equation of motion}
\end{equation}%
where $\partial_{x}=\frac{\partial}{\partial x}$ and
$\partial_{t}=\frac{\partial}{\partial t}$. Let us assume that
$U(x,t)$ is diagonizable, i.e.,
\begin{equation}
 U=g\, T \, g^{-1}, \label{constraints1}
\end{equation}%
where $g\in GL(n)$ is matrix function of $(x,t)$ and $T$ is a $n
\times n $ constant matrix
\begin{eqnarray}
T &=&\left(
\begin{array}{ccccccccc}
c_{1} & 0 & \cdots  & 0 & 0 & 0  & \cdots  & 0 & 0\\
0 & c_{1} & \cdots  & 0 & 0 & 0 & \cdots  & 0 & 0\\
\vdots  & \vdots  &   & \vdots  & \vdots & \vdots  & \vdots  & \vdots&\vdots  \\
0 & 0 & \cdots & c_{1} & 0&0 & \cdots  & 0 & 0\\
0 & 0 & \cdots & 0 &c_{2}&0 & \cdots  & 0 & 0\\
0 & 0 & \cdots & 0 &0&c_{2}& \cdots  & 0 & 0\\
\vdots  & \vdots  &    & \vdots&\vdots  & \vdots  &   & \vdots& \vdots   \\
0 & 0 & \cdots & 0 &0&0& \cdots  & 0 & c_{2}\\%
\end{array}%
\right), \label{constraints2}
\end{eqnarray}%
where $1\leq p \leq n$ and $c_{1},c_{2}\in \mathbb{R}$ (or $
\mathbb{C}$). From equations (\ref{constraints1}) and
(\ref{constraints2}), we have
\begin{equation}
\left[U,\left[U,\left[U,\chi\right]\right]\right]=c^{2}\left[U,\chi\right],
\label{constraints3}
\end{equation}%
for an arbitrary matrix function $\chi$ and $c=c_{1}-c_{2}\neq 0$.
Since
\begin{equation}
\partial_{x}U\equiv U_{x}=\left[\partial_{x}gg^{-1},U\right], \label{constraints3a}
\end{equation}%
implies
\begin{equation}
\left[U,\left[U,U_{x}\right]\right]=c^{2}U_{x},
\label{constraints4}
\end{equation}%
The equation of motion (\ref{equation of motion}) can also be
written as the zero-curvature condition i.e.,
\begin{equation}
\left[\partial_{x}-\frac{1}{(1-\lambda)}U,\partial_{t}-\frac{c^{2}}{(1-\lambda)^{2}}U-\frac{1}{(1-\lambda)}\left[U,U_{x}\right]\right]=0.
\label{zerocurvature}
\end{equation}
The above zero-curvature condition (\ref{zerocurvature}) is
equivalent to the compatibility condition of the following Lax
pair
\begin{eqnarray}
\partial_{x}\Psi(x,t;\lambda) &=&\frac{1}{(1-\lambda)} U(x,t)\Psi(x,t;\lambda)\label{linear1} \\
\partial_{t}\Psi(x,t;\lambda) &=&\left(\frac{c^{2}}{(1-\lambda)^{2}} %
U+\frac{1}{(1-\lambda)}\left[U,U_{x}\right]\right)\Psi(x,t;\lambda)\label{linear2}
\end{eqnarray}%
where $\lambda$ is a real (or complex) parameter and $\Psi$ is an
invertible $n\times n$ matrix-valued function belonging to
$GL(n)$.

In the next section, we define the Darboux transformation on
matrix solutions $\Psi$ of the Lax pair
(\ref{linear1})-(\ref{linear2}). To write down the explicit
expressions for matrix solutions of the GHM model, we will use
the notion of quasideterminant introduced by Gelfand and Retakh \cite{GR}-%
\cite{krob}.

Let $X$ be an $n\times n$ matrix over a ring $R$ (noncommutative,
in
general). For any $1\leq i$, $j\leq n$, let $r_{i}$ be the $i$th row and $%
c_{j}$ be the $j$th column of $X$. There exist $n^{2}$
quasideterminants denoted by $|X|_{ij}$ for $i,j=1,\ldots ,n$ and
are defined by
\begin{equation}
|X|_{ij}=\left\vert
\begin{array}{cc}
X^{ij} & c_{j}^{\,\,i} \\
r_{i}^{\,\,j} & \frame{\fbox{$x_{ij}$}}%
\end{array}%
\right\vert =x_{ij}-r_{i}^{\,\,j}\left( X^{ij}\right) ^{-1}c_{j}^{\,\,i},
\label{quasid}
\end{equation}%
where $x_{ij}$ is the $ij$th entry of $X$, $r_{i}^{\,\,j}$ represents the $i$%
th row of $X$ without the $j$th entry, $c_{j}^{\,\,i}$ represents the $j$th
column of $X$ without the $i$th entry and $X^{ij}$ is the submatrix of $X$
obtained by removing from $X$ the $i$th row and the $j$th column. The
quasideterminats are also denoted by the following notation. If the ring $R$
is commutative i.e. the entries of the matrix $X$ all commute, then
\begin{equation}
|X|_{ij}=(-1)^{i+j}\frac{\mathrm{det}X}{\mathrm{det}X^{ij}}.
\end{equation}%
For a detailed account of quasideterminants and their properties
see e.g. \cite{GR}-\cite{krob}. In this paper, we will consider
only quasideterminants that are expanded about an $n\times n$
matrix over a commutative ring. Let
\begin{equation}
\left(
\begin{array}{cc}
A & B \\
C & D%
\end{array}%
\right) ,  \notag
\end{equation}%
be a block decomposition of any $K\times K$ matrix where the matrix $D$ is $%
n\times n$ and $A$ is invertible. The ring $R$ in this case is the
(noncommutative) ring of $n\times n$ matrices over another
commutative ring. The quasideterminant of $K\times K$ matrix
expanded about the $n\times n$ matrix $D$ is defined by
\begin{equation}
\left\vert
\begin{array}{cc}
A & B \\
C & \frame{\fbox{$D$}}%
\end{array}%
\right\vert =D-CA^{-1}B.
\end{equation}%
The quasideterminants have found various applications in the
theory of integrable systems, where the multisoliton solutions of
various noncommutative integrable systems are expressed in terms
of quisideterminants (see e.g. \cite{Nimmo}-\cite{Hamanaka1}).

\section{Darboux transformation}

The Darboux transformation is one of the well-known method of
obtaining multi-soliton solutions of many integrable models
\cite{darboux}-\cite{matveev}. We define
the Darboux transformation on the matrix solutions of the Lax pair (\ref{linear1}%
)-(\ref{linear2}), in terms of an $n\times n$ matrix
$D(x,t,\lambda )$, called the Darboux matrix. For a general
discussion on Darboux matrix approach see e.g.
\cite{sakh}-\cite{qing ji}. The Darboux matrix relates the two
matrix solutions of the Lax pair (\ref{linear1})-(\ref{linear2}),
in such a way that the Lax pair is covariant under the Darboux
transformation. The one-fold Darboux transformation on the matrix
solution of the Lax pair (\ref{linear1})-(\ref{linear2}) is
defined by
\begin{equation}
\Psi\left[1\right](x,t;\lambda) =D(x,t,{\lambda })\Psi(x,t;\lambda
), \label{Darboux1}
\end{equation}%
where $D(x,t,{\lambda })$ is the Darboux matrix. For our case, we
can make the following ansatz
\begin{equation}
D(x,t,\lambda )=\lambda I-M(x,t),\ \ \label{Darboux2a}
\end{equation}%
where $M(x,t)$ is an $n\times n$ matrix function and $I$ is an
$n\times n$ identity matrix. The new solution
$\Psi\left[1\right](x,t;\lambda)$ satisfies the following Lax
pair, i.e.
\begin{eqnarray}
\partial_{x}\Psi\left[1\right](x,t;\lambda)  &=&\frac{1}{1-\lambda }%
U\left[1\right] \Psi\left[1\right](x,t;\lambda) ,  \label{Darboux2} \\
\partial_{t}\Psi\left[1\right](x,t;\lambda)  &=&\left(\frac{c^{2}}{(1-\lambda) ^{2}}%
 U\left[1\right]+\frac{1}{1-\lambda }%
 \left[U\left[1\right],U_{x}\left[1\right]\right]\right)\Psi\left[1\right](x,t;\lambda),  \label{Darboux3}
\end{eqnarray}%
where $U\left[1\right]$ satisfies the equation of motion
(\ref{equation of motion}). By operating $\partial_{x}$ and
$\partial_{t}$ on equation (\ref{Darboux1}) and equating the
coefficients of different powers of $\lambda$, we get the
following transformation on the matrix field $U$
\begin{eqnarray}
U\left[1\right] &=&U+ M_{x},\label{Darboux5}
\end{eqnarray}%
and the following conditions which $M$ is required to satisfy
\begin{eqnarray}
M_{x}\left(I-M\right)&=&\left[U,M\right],\label{Darboux6}\\
M_{t}\left(I-M\right)^{2}&=&\left[c^{2}U+\left[U,U_{x}\right],M\right]%
+M\left[U,U_{x}\right]M-\left[U,U_{x}\right]M^{2}.\label{Darboux7}
\end{eqnarray}%
One can solve equations (\ref{Darboux6})-(\ref{Darboux7}) to
obtain an explicit expression for the matrix function $M(x,t)$. An
explicit expression for $M(x,t)$ can be found as follows.

Let us take $n$ distinct real (or complex) constant parameters
${\lambda }_{1},\cdots ,{\lambda }_{n} (\neq 1)$. Also take $n$
constant column vectors $e_{1} ,e_{2} ,\cdots ,e_{n}$ and
construct an invertible non-degenerate $n \times n$ matrix
function $\Theta(x,t)$
\begin{equation}
\Theta(x,t)=\left( \Psi({\lambda }_{1})e_{1} ,\cdots ,\Psi({\lambda }%
_{n})e_{n}\right) =\left( \theta_{1} ,\cdots ,\theta_{n} \right).
\label{Darboux8}
\end{equation}%
Each column $\theta_{i}=\Psi({\lambda }_{i})e_{i}$ in the
matrix $\Theta$ is a column solution of the Lax pair (\ref{linear1})-(\ref{linear2}) when ${\lambda }={%
\lambda }_{i}$ and $i=1,2,\ldots ,n$ i.e.
\begin{eqnarray}
\partial_{x}\theta_{i}  &=&\frac{1}{1-\lambda _{i}}%
U\theta_{i},  \label{Darboux9} \\
\partial_{t}\theta_{i} &=&\left(\frac{c^{2}}{(1-\lambda_{i}) ^{2}}%
 U+\frac{1}{1-\lambda_{i} }%
 \left[U,U_{x}\right]\right)\theta_{i}. \label{Darboux10}
\end{eqnarray}%
Let us take an $n\times n$ invertible diagonal matrix with entries
being eigenvalues $\lambda_{i}$ corresponding to the eigenvectors
$\theta_{i}$
\begin{equation}
\Lambda =\text{diag}({\lambda }_{1},\ldots ,{\lambda }_{n}).
\label{Darboux11}
\end{equation}%
The $n\times n$ matrix generalization of the Lax pair
(\ref{Darboux9})-(\ref{Darboux10}) will be%
\begin{eqnarray}
\partial_{x}\Theta &=& U\Theta\left(I-\Lambda\right)^{-1},  \label{Darboux12} \\
\partial_{t}\Theta_{i} &=&c^{2}U\Theta\left(I-\Lambda\right)^{-2}+%
\left[U,U_{x}\right]\Theta\left(I-\Lambda\right)^{-1}.
\label{Darboux13}
\end{eqnarray}%
The $n\times n$ matrix $\Theta$ is a particular matrix solution of
the Lax pair (\ref{Darboux9})-(\ref{Darboux10}) with $\Lambda $
being a matrix of particular eigenvalues. In terms of particular
matrix solution $\Theta$ of the Lax pair
(\ref{Darboux9})-(\ref{Darboux10}), we make the following choice
of the matrix $M(x,t)$
\begin{equation}
M(x,t)=\Theta\Lambda \Theta^{-1}.  \label{Darboux14}
\end{equation}%
Our next step is to check that equation (\ref{Darboux14}) is a solution of %
equations (\ref{Darboux6})-(\ref{Darboux7}). In order to show
this, we first operate $\partial_{x}$ on equation
(\ref{Darboux14}) to get
\begin{eqnarray}
\partial _{x}M &=&\partial _{x}(\Theta\Lambda \Theta^{-1}),  \notag \\
&=&\left(\partial _{x}\Theta\right)\Lambda \Theta^{-1}+ \Theta\Lambda \partial _{x}(\Theta^{-1}),  \notag \\
&=&U\Theta(I-\Lambda )^{-1}\Lambda \Theta^{-1}-\Theta\Lambda
\Theta^{-1}U \Theta(I-\Lambda
)^{-1}\Theta^{-1},  \notag \\
&=&-U+\Theta(I-\Lambda )\Theta^{-1}j_{+}\Theta(I-\Lambda )^{-1}\Theta^{-1},  \notag \\
&=&-U+\left( I-M\right) U\left( I-M\right) ^{-1},
\label{Darboux15}
\end{eqnarray}%
which is the equation (\ref{Darboux6}). Similarly operate $\partial_{t}$ on (%
\ref{Darboux14}), we get
\begin{eqnarray}
\partial _{t}M &=&\partial _{t}\left(\Theta\Lambda \Theta^{-1}\right)  \notag \\
&=&\left(\partial _{t}\Theta\right)\Lambda \Theta^{-1}+\Theta\Theta\Lambda \partial _{t}(\Theta^{-1})  \notag \\
&=&\left(c^{2}U\Theta\left(I-\Lambda\right)^{-2}+%
\left[U,U_{x}\right]\Theta\left(I-\Lambda\right)^{-1}\right)\Lambda\Theta^{-1}-\notag\\
&&\Theta\Lambda\Theta^{-1}\left(c^{2}U\Theta\left(I-\Lambda\right)^{-2}+%
\left[U,U_{x}\right]\Theta\left(I-\Lambda\right)^{-1}\right)\Theta^{-1},
\label{Darboux17}
\end{eqnarray}%
which is equation (\ref{Darboux7}). This shows that the choice (%
\ref{Darboux14}) of the matrix $M$ satisfies the equations
(\ref{Darboux6})-(\ref{Darboux7}). In other words we can say that
if the collection $\left(\Psi,U\right)$ is a solution of the Lax
pair (\ref{linear1})-(\ref{linear2}) and the matrix $M$ is defined
by (\ref{Darboux14}), then $\left(\Psi[1],U[1]\right)$ defined by
(\ref{Darboux1}) and (\ref{Darboux5}) respectively, is also a
solution of the same Lax pair. Therefore we say that
\begin{eqnarray}
\Psi[1]&=&\left(\lambda
I-\Theta\Lambda\Theta^{-1}\right) \Psi,\notag\\
U[1]&=&\left(I-\Theta\Lambda\Theta^{-1}\right)U\left(I-\Theta\Lambda\Theta^{-1}\right)^{-1},\notag
\end{eqnarray}%
is the required Darboux transformation on the solution $\Psi$ to
the Lax pair (\ref{linear1})-(\ref{linear2}) and $U$ to the
equation of motion (\ref{equation of motion}) respectively.

\section{Quasideterminant solutions}

We have shown that the matrix $M=\Theta\Lambda\Theta^{-1}$
satisfies the conditions (\ref{Darboux6})-(\ref{Darboux7}).
Therefore, the one-fold Darboux transformation (\ref{Darboux1})
can also be written in terms of quasideterments as
\begin{eqnarray}
\Psi[1]&\equiv&D(x,t;\lambda)\Psi=\left(\lambda
I-\Theta_{1}\Lambda_{1}\Theta_{1}^{-1}\right) \Psi,\notag \\
&=&\left\vert
\begin{array}{cc}
\Theta_{1} & \Psi \\
\Theta_{1}\Lambda_{1} & \frame{\fbox{$\lambda \Psi$}}%
\end{array}%
\right\vert. \label{solution1}
\end{eqnarray}%
The above equation defines the Darboux transformation on the
matrix solution $\Psi$ of the Lax pair
(\ref{linear1})-(\ref{linear2}). The corresponding one-fold
Darboux transformation on the matrix field $U$ is
\begin{eqnarray}
U[1]&=&\left(I-\Theta_{1}\Lambda_{1}\Theta_{1}^{-1}\right)U\left(I-\Theta_{1}\Lambda_{1}\Theta_{1}^{-1}\right)^{-1},\notag\\
&=&\left\vert
\begin{array}{cc}
\Theta_{1} & I \\
\Theta_{1}\left(I-\Lambda_{1}\right) & \frame{\fbox{$0$}}%
\end{array}%
\right\vert U \left\vert
\begin{array}{cc}
\Theta _{1}& I \\
\Theta_{1}\left(I-\Lambda_{1}\right) & \frame{\fbox{$0$}}%
\end{array}%
\right\vert^{-1} .\label{solution2}
\end{eqnarray}%
We write two-fold Darboux transformation on $\Psi$ as%
\begin{eqnarray}
\Psi[2]&\equiv&D(x,t;\lambda)\Psi[1]= \lambda\Psi[1]-\Theta_{2}[1]\Lambda_{2}\Theta^{-1}_{2}[1]\Psi[1]\notag\\
&=&\lambda\left(\lambda
I-\Theta_{1}\Lambda_{1}\Theta_{1}^{-1}\right) \Psi-\notag\\
&&\left(\Theta_{2}\Lambda_{2}-\Theta_{1}\Lambda_{1}\Theta^{-1}_{1}\Theta_{2}\right)%
\Lambda_{2}\left(\Theta_{2}\Lambda_{2}-\Theta_{1}\Lambda_{1}\Theta^{-1}_{1}\Theta_{2}\right)^{-1}\left(\lambda
I-\Theta_{1}\Lambda_{1}\Theta_{1}^{-1}\right) \Psi,\notag \\
&=&\left\vert
\begin{array}{ccc}
\Theta_{1}& \Theta_{2} & \Psi \\
\Theta_{1}\Lambda_{1} &\Theta_{2}\Lambda_{2} & \lambda\Psi\\
\Theta_{1}\Lambda_{1}^{2}&\Theta_{2}\Lambda_{2}^{2} & \frame{\fbox{$\lambda^{2} \Psi$}}%
\end{array}%
\right\vert. \label{solution3}
\end{eqnarray}%
Similarly the expression for two-fold Darboux transformation on
the matrix field $U$ as
\begin{eqnarray}
U[2]&=& \Theta_{2}[1]\left(I-\Lambda_{2}\right)\Theta^{-1}_{2}[1]U[1]%
\left(\Theta_{2}[1]\left(I-\Lambda_{2}\right)\Theta^{-1}_{2}[1]\right)^{-1},\notag\\
&=& \left(\Theta_{2}\Lambda_{2}-\Theta_{1}\Lambda_{1}\Theta^{-1}_{1}\Theta_{2}\right)%
\left(I-\Lambda_{2}\right)\left(\Theta_{2}\Lambda_{2}-\Theta_{1}\Lambda_{1}\Theta^{-1}_{1}\Theta_{2}\right)^{-1}\times\notag\\
&&\left(I-\Theta_{1}\Lambda_{1}\Theta_{1}^{-1}\right)U\left(I-\Theta_{1}\Lambda_{1}\Theta_{1}^{-1}\right)^{-1}\times\notag\\
&&\left(\left(\Theta_{2}\Lambda_{2}-\Theta_{1}\Lambda_{1}\Theta^{-1}_{1}\Theta_{2}\right)%
\left(I-\Lambda_{2}\right)\left(\Theta_{2}\Lambda_{2}-\Theta_{1}\Lambda_{1}\Theta^{-1}_{1}\Theta_{2}\right)^{-1}\right)^{-1},\notag%
\end{eqnarray}
\begin{eqnarray}
 &=&\left\vert\begin{array}{ccc}
\Theta_{1}& \Theta_{2} & I \\
\Theta_{1}\left(I-\Lambda_{1}\right) &\Theta_{2}\left(I-\Lambda_{2}\right) & 0\\
\Theta_{1}\left(I-\Lambda_{1}\right)^{2}&\Theta_{2}\left(I-\Lambda_{2}\right)^{2} & \frame{\fbox{$0$}}%
\end{array}%
\right\vert\times U\times\notag\\
&&\times\left\vert
\begin{array}{ccc}
\Theta_{1}& \Theta_{2} & I \\
\Theta_{1}\left(I-\Lambda_{1}\right) &\Theta_{2}\left(I-\Lambda_{2}\right) & 0\\
\Theta_{1}\left(I-\Lambda_{1}\right)^{2}&\Theta_{2}\left(I-\Lambda_{2}\right)^{2} & \frame{\fbox{$0$}}%
\end{array}%
\right\vert^{-1}. \label{solution4}
\end{eqnarray}%

The result can be generalized to obtain $N$-fold Darboux
transformation on matrix solution $\Psi$ as
\begin{eqnarray}
\Psi[N] &=&\left\vert
\begin{array}{ccccc}
\Theta_{1} & \Theta_{2} & \cdots  & \Theta_{N} & \Psi \\
\Theta_{1}\Lambda _{1} & \Theta_{2}\Lambda _{2} & \cdots  &
\Theta_{N}\Lambda _{N} &
\lambda \Psi \\
\Theta_{1}\Lambda _{1}^{2} & \Theta_{2}\Lambda _{2}^{2} & \cdots &
\Theta_{N}\Lambda
_{N}^{2} & \lambda ^{2}\Psi \\
\vdots  & \vdots  & \ddots  & \vdots  & \vdots  \\
\Theta_{1}\Lambda _{1}^{N} & \Theta_{2}\Lambda _{2}^{N} & \cdots &
\Theta_{N}\Lambda
_{N}^{N} & \frame{\fbox{$\lambda ^{N}\Psi$}}%
\end{array}%
\right\vert .  \label{solution5}
\end{eqnarray}%
Similarly the expression for $U[N]$ is
\begin{eqnarray}
U[N]&=&\left\vert
\begin{array}{ccccc}
\Theta_{1}& \Theta_{2} & \cdots &\Theta_{N}& I \\
\Theta_{1}\left(I-\Lambda_{1}\right) &\Theta_{2}\left(I-\Lambda_{2}\right) & \cdots &\Theta_{N}\left(I-\Lambda_{N}\right)&  0\\
\Theta_{1}\left(I-\Lambda_{1}\right)^{2}&\Theta_{2}\left(I-\Lambda_{2}\right)^{2}
& \cdots&\Theta_{N}\left(I-\Lambda_{N}\right)^{2} &  0\\%
\vdots  & \vdots  & \ddots  & \vdots  & \vdots  \\
\Theta_{1}\left(I-\Lambda_{1}\right)^{N}&\Theta_{2}\left(I-\Lambda_{2}\right)^{N}
& \cdots &
\Theta_{N}\left(I-\Lambda_{N}\right)^{N}&\frame{\fbox{$0$}}
\end{array}%
\right\vert\times U\times\notag\\
&&\times\left\vert
\begin{array}{ccccc}
\Theta_{1}& \Theta_{2} & \cdots &\Theta_{N}& I \\
\Theta_{1}\left(I-\Lambda_{1}\right) &\Theta_{2}\left(I-\Lambda_{2}\right) & \cdots &\Theta_{N}\left(I-\Lambda_{N}\right)&  0\\
\Theta_{1}\left(I-\Lambda_{1}\right)^{2}&\Theta_{2}\left(I-\Lambda_{2}\right)^{2}
& \cdots&\Theta_{N}\left(I-\Lambda_{N}\right)^{2} &  0\\%
\vdots  & \vdots  & \ddots  & \vdots  & \vdots  \\
\Theta_{1}\left(I-\Lambda_{1}\right)^{N}&\Theta_{2}\left(I-\Lambda_{2}\right)^{N}
& \cdots &
\Theta_{N}\left(I-\Lambda_{N}\right)^{N}&\frame{\fbox{$0$}}
\end{array}%
\right\vert^{-1}. \label{solution4a}
\end{eqnarray}%
We now relate the quasideterminant solutions of GHM with the
solutions obtained by dressing method and the inverse scattering
method. For this purpose, we proceed  as follows. From the
definition of the matrix $M$, we have
\begin{eqnarray}
M\Theta&=&\Theta\Lambda.\label{solution5a}
\end{eqnarray}
Let $\theta_{i}$ and $\theta_{j}$ be the column solutions of the
Lax pair (\ref{linear1})-(\ref{linear2}) when
$\lambda=\lambda_{i}$ and $\lambda=\lambda_{j}$ respectively i.e.
\begin{eqnarray}
M\theta_{i}&=&\lambda_{i}\theta_{i},\quad i=1,2,\dots, p\notag\\
M\theta_{j}&=&\lambda_{j}\theta_{j}.\quad j=p+1,p+2,\dots, n
\label{solution6}
\end{eqnarray}
Now we take $\lambda_{i}=\mu$ and $\lambda_{j}=\bar{\mu}$, we may
write the matrix $M$ as
\begin{eqnarray}
M&=&\mu P + \bar{\mu}P^{\perp}, \label{solution7}
\end{eqnarray}
where $P$ is the hermitian projector i.e. $P^{\dagger}=P$. The
projector $P$ satisfies $P^2=P$ and $P^{\perp}=1-P$. The projector
$P$ is hermitian projection on a complex space and $P^{\perp}$ as
projection on orthogonal space. Now equation (\ref{solution7}) can
also written as
\begin{eqnarray}
M&=&\left(\mu-\bar{\mu}\right) P + \bar{\mu}I, \label{solution8}
\end{eqnarray}
where the hermitian projector can be expressed as
\begin{eqnarray}
P&=&\theta_{i}\left(\theta_{i}^\dagger
,\theta_{i}\right)^{-1}\theta_{i}^\dagger. \label{solution8a}
\end{eqnarray}
The one-fold Darboux transformation (\ref{solution1}) on the
matrix solution $\Psi$ can also be expressed in terms of projector
$P$ as
\begin{eqnarray}
\Psi[1]&\equiv&{\cal
D}(x,t;\lambda)\Psi=\left(I-\frac{\mu-\bar{\mu}}{\lambda-\bar{\mu}}P\right)\Psi,\label{solution9}
\end{eqnarray}
where ${\cal D}(x,t;\lambda)$ is the rescaled Darboux-dressing
function i.e. ${\cal
D}(x,t;\lambda)=\left(\lambda-\mu\right)^{-1}D(x,t;\lambda)$.
Similarly the $N$-fold Darboux transformation (\ref{solution5}) on
the matrix solution $\Psi$ can also be written as (take $P[1]=P$)
\begin{eqnarray}
\Psi[N]&=&\prod_{k=0}^{N-1}%
\left(I-\frac{\mu_{N-k}-{\bar{\mu}_{N-k}}}{\lambda-{\bar{\mu}_{N-k}}}P[N-k]\right)\Psi.\label{solution11}
\end{eqnarray}
Now we can express the $N$-fold Darboux transformation
(\ref{solution4a}) on the matrix field $U$ can be written as
\begin{eqnarray}
U[N]&=&\prod_{k=0}^{N-1}%
\left(I-\frac{\mu_{N-k}-{\bar{\mu}_{N-k}}}{1-{\bar{\mu}_{N-k}}}P[N-k]\right)%
U\prod_{l=1}^{N-1}\left(I-\frac{{\bar{\mu}_{l}}-\mu_{l}}{1-{\bar{\mu}_{l}}}P[l]\right),\label{solution12}
\end{eqnarray}
and hermitian projector is defined as
\begin{eqnarray}
P[k]&=&\theta_{i}[k]\left(\theta_{i}^{\dagger}[k],\theta_{i}[k]\right)^{-1}\theta_{i}^{\dagger}[k].
\label{solution8aa}
\end{eqnarray}
The expressions (\ref{solution11}) and (\ref{solution12}) can also
be written as sum of $K$ terms \cite{U(N)}
\begin{eqnarray}
\Psi[N]&=&\sum_{k=0}^{N-1}\left(I-\frac{1}{\lambda-{\bar{\mu}_{k}}}R_{k}\right)\Psi,\label{solution13}
\end{eqnarray}
and
\begin{eqnarray}
U[N]&=&\sum_{k=0}^{N-1}%
\left(I-\frac{1}{1-{\bar{\mu}_{k}}}R_{k}\right)%
U\sum_{l=0}^{N-1}\left(I-\frac{1}{1-{\bar{\mu}_{l}}}R_{l}\right)^{-1},\label{solution14}
\end{eqnarray}
where
\begin{eqnarray}
R_{k}&=&\sum_{l=0}^{N-1}%
\left(\mu_{l}-\bar{\mu_{k}}\right)%
\theta_{i}^{(k)}\left(\theta_{i}^{(k)\dagger},\theta_{i}^{(l)}\right)^{-1}\theta_{i}^{(l)\dagger}.\label{solution15}
\end{eqnarray}

\section{The explicit solutions of the GHM model}

In this section we calculate explicit expression of soliton
solution. First of all we will study GHM model based on $SU(n)$.
In this case the spin function $U$ takes values in the Lie algebra
$\verb"su(n)"$ so that one can decompose the spin function into
components $U=U^{a}T^{a}$, and $T^{a}, a=1,2,\dots,n^{2}$ are
anti-hermitian $n\times n$ matrices with normalization
$\mbox{Tr}\left(T^{a}T^{b}\right)=\frac{1}{2}\delta^{ab}$ and are
the generators of the $SU(n)$ in the fundamental representation
satisfying the algebra
\begin{equation}
\left[T^{a},T^{b}\right]=f^{abc}T^{c}, \label{action1}
\end{equation}%
where $f^{abc}$ are the structure constants of the Lie algebra
$\verb"su(n)"$. For any $X \in \verb"su(n)"$, we write
$X=X^{a}T^{a}$ and $U^{a}=-2\mbox{Tr}(UT^{a})$.

The matrix-field $U$ belongs to the Lie algebra $\verb"su(n)"$ of
the Lie group $SU(n)$ therefore
\begin{eqnarray}
U^\dagger=-U, \quad\quad \mbox{Tr}(U)=0. \label{su1}
\end{eqnarray}
The equations (\ref{Darboux1})-(\ref{Darboux2a}) and
(\ref{Darboux5}) define a Darboux transformation for the GHM model
based on the Lie group $SU(n)$. The new solution of the equation
of motion (\ref{equation of motion}) $U[1]$ must be $\verb"su(n)"$
valued i.e.
\begin{eqnarray}
U^\dagger[1]=-U[1], \quad\quad \mbox{Tr}(U[1])=0, \label{su2}
\end{eqnarray}
therefore, we have the following conditions on the matrix $M$
\begin{eqnarray}
M^\dagger=-M, \quad\quad \mbox{Tr}(M)=0. \label{su3}
\end{eqnarray}
In other words we want to make specific $M$ to satisfy the
(\ref{su3}). This can be achieved if we choose the particular
solutions $\theta_{i}$ at $\lambda=\lambda_{i}$, let us first
calculate
\begin{eqnarray}
\partial_{x}\left(\theta_{i}^{\dagger}\theta_{j}\right)&=&%
\left(\partial_{x}\theta_{i}^{\dagger}\right)\theta_{j}+%
\theta_{i}^{\dagger}\left(\partial_{x}\theta_{j}\right)\notag\\%
&=&\left(1-\bar{\lambda_{i}}\right)^{-1}\theta_{i}^{\dagger}U^{\dagger}\theta_{j}+%
\left(1-\lambda_{j}\right)^{-1}\theta_{i}^{\dagger}U\theta_{j},\label{su4}
\end{eqnarray}
using equation (\ref{su1}) the above equation (\ref{su4}) becomes
\begin{eqnarray}
\partial_{x}\left(\theta_{i}^{\dagger}\theta_{j}\right)&=&0,\label{su5}
\end{eqnarray}
when $\lambda_{i}\neq\lambda_{j}$ (i.e.
$\bar{\lambda_{i}}=\lambda_{j}$). Similarly we can check
\begin{eqnarray}
\partial_{t}\left(\theta_{i}^{\dagger}\theta_{j}\right)&=&0.\label{su6}
\end{eqnarray}
From the definition of the matrix $M$, we have
\begin{eqnarray}
\theta_{i}^{\dagger}\left(M^{\dagger}+M\right)\theta_{j}&=&%
\left(\bar{\lambda_{i}}+\lambda_{j}\right)\theta_{i}^{\dagger}\theta_{j},\label{su7}
\end{eqnarray}
when $\lambda_{i}\neq\lambda_{j}$ then the above expression
(\ref{su7}) implies
\begin{eqnarray}
\theta_{i}^{\dagger}\theta_{j}=0.\label{su7a}
\end{eqnarray}
The column vectors $\theta_{i}$ are linearly independent and the
equation (\ref{su7a}) holds everywhere.

For the HM model based on $SU(n)$, the constant matrix
(\ref{constraints2}) becomes
\begin{eqnarray}
T &=&\left(
\begin{array}{ccccccccc}
2-\frac{2}{n} & 0 & \cdots  & 0 & 0 & 0  & \cdots  & 0 & 0\\
0 & -\frac{2}{n} & \cdots  & 0 & 0 & 0 & \cdots  & 0 & 0\\
\vdots  & \vdots  &   & \vdots  & \vdots & \vdots  & \vdots  & \vdots&\vdots  \\
0 & 0 & \cdots & -\frac{2}{n} & 0&0 & \cdots  & 0 & 0\\
0 & 0 & \cdots & 0 &-\frac{2}{n}&0 & \cdots  & 0 & 0\\
0 & 0 & \cdots & 0 &0&-\frac{2}{n}& \cdots  & 0 & 0\\
\vdots  & \vdots  &    & \vdots&\vdots  & \vdots  &   & \vdots& \vdots   \\
0 & 0 & \cdots & 0 &0&0& \cdots  & 0 &-\frac{2}{n}\\%
\end{array}%
\right). \label{constraints1a11}
\end{eqnarray}%
Then $U^{2}$ becomes
\begin{equation}
U^{2}=\frac{4\left(n-1\right)}{n^{2}}I+\frac{2\left(n-2\right)}{n}U.\label{constraints211}
\end{equation}%
These are the constraints given in ref. \cite{GHM3}. For the
construction of explicit soliton solution for the $SU(n)$ HM
model, we construct the matrix $M$ by defining a Hermitian
projector $P$. For this case, we take the seed solution to be
\begin{equation}
U_{0}\equiv U=\mbox{i}\left(
\begin{array}{ccc}
a_{1} &  &  \\
& \ddots  &  \\
&  & a_{n}
\end{array}%
\right),
\end{equation}
where $a_i $ are real constants and $\sum_{i=1}^{n} a_i =0 $. The
corresponding solution of the Lax pair is expressed in block
diagonal matrix
\begin{equation}
\Psi (x,t;\lambda )=\left(
\begin{array}{cc}
W_{p}(\lambda ) & O \\
O & W_{n-p}(\lambda )%
\end{array}%
\right),
\end{equation}
where
\begin{equation} W_{p}(\lambda )=\left(
\begin{array}{ccc}
e^{\mbox{i}\omega _{1}(\lambda )} &  &  \\
& \ddots  &  \\
&  & e^{\mbox{i}\omega _{p}(\lambda )}%
\end{array}%
\right),
\end{equation}
and
\begin{equation} W_{n-p}(\lambda )=\left(
\begin{array}{ccc}
e^{\mbox{i}\omega _{p+1}(\lambda )} &  &  \\
& \ddots  &  \\
&  & e^{\mbox{i}\omega _{n}(\lambda )}%
\end{array}%
\right),
\end{equation}
are $p\times p$ and $(n-p)\times(n-p) $ matrices respectively and
\begin{equation} \omega _{i}(\lambda )=a_{i}\left( \frac{1}{1-\lambda
}x+\frac{4}{\left( 1-\lambda \right) ^{2}}t\right).
\end{equation}
Now define a particular matrix solution $\Theta $ of the Lax pair as
\begin{equation}
\Theta =\left( \Psi (\mu )L_{1}\ ,\  \Psi (\bar{\mu})L_{2}\right),
\end{equation}
where $L_1 $ is an $n \times p$ constant matrix of $p$ column
vectors and $L_2 $ is the orthogonal complementary $n \times
(n-p)$ matrix of $(n-p)$ column vectors. The columns of $L_1$ span
a $p$-dimensional subspace $U$ of $C^n $, and those of $L_2 $ span
the orthogonal subspace $V$. The projector $P$ is completely
characterized by the two subspaces $U=\text{Im} P $ and
$V=\text{Ker} P $ given by the condition $P^\bot U=0 $ and $PV=0
$. Let us write $ L_{1}=\left(
\begin{array}{c}
A \\
B%
\end{array}%
\right) $ and $ L_{2}=\left(
\begin{array}{c}
C \\
D%
\end{array}%
\right), $ where $A$, $B$, $C$ and $D$ are constant $p \times p$,
$(n-p) \times n $, $p \times (n-p)$ and $(n-p) \times (n-p)$
constant matrices respectively. Given this, the $n \times n$ matric
$\Theta $ is given by
\begin{equation} \Theta =\left(
\begin{array}{cc}
W_{p}(\mu )A & W_{p}(\bar{\mu})C \\
W_{n-p}(\mu )B & W_{n-p}(\bar{\mu})D%
\end{array}%
\right).
\end{equation}
We now define the projector $P$ in terms of the matrix
$\Phi=\Psi(\mu ) L_1 =(\theta_1 , \cdots \theta_p ) $ given by
\begin{eqnarray*}
\Phi  &=&\left( \theta _{1},\cdots ,\theta _{p}\right)  \\
&=&\left(
\begin{array}{c}
W_{p}(\mu )A \\
W_{n-p}(\mu )B%
\end{array}%
\right).
\end{eqnarray*}
The projector is thus given by
\begin{equation}
P=\left(
\begin{array}{cc}
W_{p}(\mu )A\Delta A^{\dag }W_{p}^{\dag }(\bar{\mu}) & W_{p}(\mu
)A\Delta
B^{\dag }W_{n-p}^{\dag }(\bar{\mu}) \\
W_{n-p}(\mu )B\Delta A^{\dag }W_{p}^{\dag }(\bar{\mu}) & W_{n-p}(\mu
)B\Delta B^{\dag }W_{n-p}^{\dag }(\bar{\mu})%
\end{array}%
\right),
\end{equation}
where $ \Delta ^{-1}=A^{\dag }W_{p}^{\dag }(\bar{\mu})W_{p}(\mu
)A+B^{\dag }W_{n-p}^{\dag }(\bar{\mu})W_{n-p}(\mu )A $. The
Darboux matrix $D(\lambda )$ can now be constructed to give
explicit soliton solution of the $SU(n)$ HM model. To elaborate
the result more explicitly, we proceed with the example of $SU(2)$
HM model.

For the $SU(2)$ model, the equations (\ref{constraints1a11}) and
(\ref{constraints211}) become
\begin{eqnarray}
T=\left(\begin{array}{cc}
1 & 0 \\
0 & -1 \\
\end{array}%
\right). \label{constraints1a11a}
\end{eqnarray}%
Then $U^{2}$ becomes
\begin{equation}
U^{2}=I.\label{constraints211a}
\end{equation}
The Lax pair (\ref{linear1})-(\ref{linear2}) for the $SU(2)$ model
can be written as
\begin{eqnarray}
\partial_{x}\Psi(x,t;\lambda) &=&\frac{1}{(1-\lambda)} U(x,t)\Psi(x,t;\lambda),\label{linear1a} \\
\partial_{t}\Psi(x,t;\lambda) &=&\left(\frac{4}{(1-\lambda)^{2}} %
U+\frac{2}{(1-\lambda)}UU_{x}\right)\Psi(x,t;\lambda).\label{linear2a}
\end{eqnarray}%
If we take trivial solution (as seed solution), single soliton and
multi-soliton solutions can be obtained by Darboux transformation
as explained above.

We take the seed solution to be
\begin{eqnarray}
U_{0}\equiv U=\left(
\begin{array}{cc}
\mbox{i}&0\\
0 &-\mbox{i}
\end{array}
\right).\label{su8}
\end{eqnarray}
The corresponding solution of the linear system
(\ref{linear1a})-(\ref{linear2a}) can be written as
\begin{eqnarray}
\Psi(x,t;\lambda)=\left(
\begin{array}{cc}
e^{\mbox{i}\left(\frac{1}{\left(1-\lambda\right)}x+\frac{4}{\left(1-\lambda\right)^{2}}t\right)}&0\\
0
&e^{-\mbox{i}\left(\frac{1}{\left(1-\lambda\right)}x+\frac{4}{\left(1-\lambda\right)^{2}}t\right)}
\end{array}
\right).\label{su9}
\end{eqnarray}
Take $\lambda_{1}=\mu$ and $\lambda_{2}=\bar{\mu}$, the constant
matrix $\Lambda$ is given by
\begin{eqnarray}
\Lambda=\left(
\begin{array}{cc}
\mu & 0\\
0 & \bar{\mu}
\end{array}
\right),\label{su9a}
\end{eqnarray}
and corresponding $2 \times 2$ matrix solution $\Theta$ becomes
\begin{eqnarray}
\Theta\equiv\left(\theta_{1},\theta_{2}\right)=\left(
\begin{array}{cc}
e^{\mbox{i}\left(\frac{1}{\left(1-\mu\right)}x+\frac{4}{\left(1-\mu\right)^{2}}t\right)}&e^{\mbox{i}\left(\frac{1}{\left(1-\bar{\mu}\right)}x+\frac{4}{\left(1-\bar{\mu}\right)^{2}}t\right)}\\
-e^{-\mbox{i}\left(\frac{1}{\left(1-\mu\right)}x+\frac{4}{\left(1-\mu\right)^{2}}t\right)}
&e^{-\mbox{i}\left(\frac{1}{\left(1-\bar{\mu}\right)}x+\frac{4}{\left(1-\bar{\mu}\right)^{2}}t\right)}
\end{array}
\right).\label{su9c}
\end{eqnarray}

The matrix $M$ is given by
\begin{eqnarray}
M &=&\Theta\Lambda \Theta^{-1},  \notag \\
&=&\frac{1}{e^{u}+e^{-u}}\left(
\begin{array}{ll}
\mu e^{u}+{\bar{\mu}}{e^{-u}} & \left( {\bar{\mu}-\mu }\right)
e^{{i}v}
\\
\left( {\bar{\mu}-\mu }\right) e^{-iv} & {\bar{\mu}}{e^{u}}+{\mu}{e^{-u}}%
\end{array}%
\right) ,  \label{su14}
\end{eqnarray}%
where the functions $u(x,t)$ and $v(x,t)$ are defined by
\begin{eqnarray}
u(x,t) &=&\mbox{i}\left( \frac{1}{\left( 1-\mu \right)
}-\frac{1}{\left(
1-\bar{\mu}\right) }\right) x+4\mbox{i}\left( \frac{1}{\left( 1-\mu \right)^{2} }-%
\frac{1}{\left( 1-\bar{\mu}\right)^{2} }\right) t,  \notag \\
v(x,t) &=&\left( \frac{1}{\left( 1-\mu\right) }+\frac{1}{\left( 1-%
\bar{\mu}\right) }\right) x+4\left( \frac{1}{\left( 1-\mu\right)^{2} }+\frac{1}{\left( 1-%
\bar{\mu}\right) ^{2}}\right) t.  \label{rs}
\end{eqnarray}
Let us take the eigenvalue to be $\mu =e^{\mbox{i}\theta }.$ The expression (\ref%
{su14}) then becomes%
\begin{equation}
M=\left(
\begin{array}{cc}
\cos \theta +\mbox{i}\sin \theta \tanh u & -\mbox{i}\left( \sin
\theta \text{sech}u\right)
e^{\mbox{i}v} \\
-\mbox{i}\left( \sin \theta \text{sech}u\right) e^{-\mbox{i}v} &
\cos \theta -\mbox{i}\sin \theta
\tanh u%
\end{array}%
\right) ,
\end{equation}%
and the corresponding Darboux matrix $D\left( \lambda \right) $ in
this case is
\begin{equation}
D\left( \lambda \right) =\left(
\begin{array}{cc}
\lambda -\cos \theta -\mbox{i}\sin \theta \tanh u & \mbox{i}\left( \sin \theta \text{sech}%
u\right) e^{\mbox{i}v} \\
\mbox{i}\left( \sin \theta \text{sech}u\right) e^{-\mbox{i}v} &
\lambda -\cos \theta
+\mbox{i}\sin \theta \tanh u%
\end{array}%
\right) .
\end{equation}%
Comparing the above equation with (\ref{solution9}), we find the
following
expression for the projector%
\begin{equation}
P=\left(
\begin{array}{cc}
2e^{u}\text{sech}u & -2e^{\mbox{i}v}\text{sech}u \\
-2e^{-\mbox{i}v}\text{sech}u & 2e^{-u}\text{sech}u%
\end{array}%
\right).
\end{equation}%
Using (\ref{solution2}) and (\ref{su8}), we get
\begin{equation}
U[1]=\left(
\begin{array}{cc}
 \mbox{i} U_{3}& U_{+}\\
 -U_{-}&-\mbox{i} U_{3}\\%
\end{array}\right),
\end{equation}
where
\begin{eqnarray}
U_{3}&=&1-(1+\cos\theta)\mbox{sech}^{2}u ,\notag\\
U_{+}&\equiv&\overline{U}_{-}=-\mbox{i}e^{\mbox{i}v}\left[(1+\cos\theta)\mbox{tanh}u+\mbox{i}\sin\theta\right]\mbox{sech}u.\label{s123}
\end{eqnarray}
From equation (\ref{s123}), we see that $U^{\dagger}[1]=-U[1]$ and
$\mbox{Tr}(U[1])=0$. Therefore equation (\ref{s123}) is an
explicit expression of the single-soliton solution of the HM model
based on $SU(2)$ obtained by using Darboux transformation.
Similarly one can calculate explicit expression for the
multi-soliton solution of the model. The expression (\ref{s123})
is similar to the expression of the single soliton given in
\cite{GHM1}.
\section{Concluding remarks}

In this paper, we have studied GHM model based on general linear
Lie group $GL(n)$ and expressed the multi-soliton solutions in
terms of the quasideterminant using the Darboux transformation
defined on the solution of the Lax pair. We have also established
equivalence between the Darboux matrix approach and the
Zakharov-Mikhailov's dressing method. In last section we have
reduced the GHM model into the HM model based on $SU(n)$ and
calculated an explicit expression for the single-soliton solution.
It would be interesting to study the GHM models based on Hermitian
symmetric spaces. We shall address this problem in a separate
work.


\end{document}